\pdfoutput=1

\documentclass[
  aps,
  pra,
  reprint,
  amsmath,
  amssymb,
  superscriptaddress,
  longbibliography,
  floatfix
]{revtex4-2}

\usepackage{graphicx}
\usepackage{dcolumn}
\usepackage{bm}
\usepackage{mathtools}
\usepackage{microtype}
\usepackage[colorlinks=true,allcolors=blue]{hyperref}

\newcommand{\PR}{\mathrm{PR}}
\newcommand{\DP}{\mathcal{D}_{\mathcal{P}}}
\newcommand{\Seq}{\mathcal{S}_{\beta}}
\newcommand{\rall}{\overline{r}_{\mathrm{all}}}
\newcommand{\ADM}{\mathrm{D}}
\newcommand{\ARM}{\mathrm{R}}

\begin{document}

\title{Chaos and scrambling in anisotropic Dicke and Rabi models}

\author{Long-Jie Li}

\author{Li-Li Gao}

\author{Zi-Cheng Liu}

\affiliation{Institute of Quantum Physics, Hunan Key Laboratory of Nanophotonics and Devices, and Hunan Key Laboratory of Super-Microstructure and Ultrafast Process, School of Physics, Central South University, Changsha 410083, China}

\author{Zi-Min Li}
\email{zimin.li@csu.edu.cn}
\affiliation{Institute of Quantum Physics, Hunan Key Laboratory of Nanophotonics and Devices, and Hunan Key Laboratory of Super-Microstructure and Ultrafast Process, School of Physics, Central South University, Changsha 410083, China}

\author{Murray T. Batchelor}
\affiliation{Mathematical Sciences Institute, Australian National University, Canberra ACT 2601, Australia}

\date{October 20, 2026}

\begin{abstract}

Identifying the quantum signatures of classical chaos remains a central question in quantum physics, and quantum scrambling provides a new perspective on this correspondence.
We study two distinct limits of the same model to examine how the correspondence between chaos and scrambling changes across different regimes.
The thermodynamic-limit anisotropic Dicke model exhibits a close correspondence between chaos and scrambling, whereas this correspondence breaks down in the large-frequency-ratio limit of the anisotropic Rabi model (ARM).
The contrasting chaotic behavior in the two limits originates from the different numbers of classical degrees of freedom, whereas their similar scrambling behavior stems from eigenstate delocalization.
This implies that, in the large-frequency-ratio limit, eigenstate delocalization alone is insufficient to qualify the ARM as a classically chaotic model.
This finding suggests that identifying chaos in quantum systems requires jointly considering the classical degrees of freedom and eigenstate delocalization.

\end{abstract}


\maketitle

\section{Introduction}
\label{sec:intro}

Understanding how classical chaos manifests in quantum systems remains a central problem in quantum physics.
Quantum scrambling provides a complementary perspective on this classical--quantum correspondence.
Light--matter interaction systems offer a versatile platform for exploring the question, with strong coupling, broad tunability, and direct access to nonequilibrium dynamics~\cite{FornDiaz2010,Yoshihara2017,Blais2021,Dimer2007,
Nagy2010,Baumann2010,Puebla2017,Cai2021}.

In these systems, whether the superradiant phase transition induces quantum chaos has long been an open question.
Early studies of the Dicke model found a Poisson-to-Wigner--Dyson crossover in the critical coupling region and related it to precursors of the ground-state transition~\cite{Kus1985,EmaryBrandes2003PRL,EmaryBrandes2003PRE}.
Energy-resolved studies subsequently examined the relation between the onset of chaos and excited-state quantum phase transitions (ESQPTs), while also showing that the two have distinct underlying mechanisms~\cite{PerezFernandez2011,ChavezCarlos2016}.
Subsequent studies of the anisotropic Dicke model (ADM) showed that the chaos crossover is more closely tied to the departure from the integrable axes than to the superradiant phase transition itself~\cite{Buijsman2017,Das2023}.

This question has recently resurfaced in the Rabi model, where the fidelity out-of-time-order correlator (FOTOC) exhibits rapid growth in the superradiant phase and has been interpreted as a signature of chaos~\cite{Kirkova2022}.
However, related studies have shown that OTOC growth can capture quantum scrambling without necessarily providing a definitive diagnostic of quantum chaos~\cite{XuScaffidiCao2020,Pilatowsky2020,Rozenbaum2020,Hashimoto2020}.
Wang \emph{et al.}~\cite{Wang2025} attributed early-time OTOC growth in the anisotropic Rabi model (ARM), including for classically stable initial conditions, to quantum collapse of the evolving wave packet, while our companion Letter~\cite{Companion} further relates scrambling to the local instability of saddle points.

In the thermodynamic limit, the collective spin of the ADM survives as an independent canonical pair.
By contrast, in the large-frequency-ratio limit, the ARM is reduced to two adiabatic branches associated with the single qubit, with the dynamics confined to an individual branch.
Despite this difference, the ADM and ARM exhibit equivalent ground-state superradiant phase transitions in their respective limits.
Following the historical development of the Dicke model, we investigate the ARM from the complementary perspectives of chaos and scrambling and compare its behavior with that of the ADM.

We construct a family of adiabatic energy surfaces for the two limits to characterize their underlying static energy-landscape topology.
The ADM and ARM share the same family of adiabatic energy surfaces and therefore possess the same ground-state properties and stationary-point structure.
In the thermodynamic limit, however, the ADM retains a continuous branch structure together with an independent collective-spin canonical pair, yielding two classical degrees of freedom, whereas the large-frequency-ratio ARM reduces to only one.
We characterize this difference quantitatively through the density of states and qualitatively through the semiclassical dynamics.

Against this contrast, and taking the ADM as a reference, we use spectral statistics and thermal OTOCs as diagnostics of chaos and scrambling, respectively.
The spectral statistics show that the ADM exhibits the standard GOE distribution in the nonintegrable region, whereas the ARM displays no analogous random-matrix level correlations.
The OTOC analysis, however, reveals an enhanced infinite-time plateau in the nonintegrable region of the ADM and a plateau of a comparable magnitude in the ARM.
By evaluating the thermally weighted participation ratio (PR) of the eigenstates in the bare parity basis as a measure of delocalization, we closely track the OTOC coupling-plane maps of both the ADM and the ARM.
These results suggest that identifying chaos in quantum systems requires jointly considering the number of classical degrees of freedom and eigenstate delocalization.

The rest of this paper is organized as follows.
Section~\ref{sec:models} introduces the models, their adiabatic energy surfaces, and the corresponding density-of-states analysis.
Section~\ref{sec:semiclassical} presents the semiclassical dynamics of the ADM and ARM through Poincar\'e sections.
Sections~\ref{sec:spectral} and~\ref{sec:scrambling} diagnose chaos and scrambling in the ADM and ARM using spectral statistics and thermal OTOCs, respectively.
Section~\ref{sec:deloc} quantifies eigenstate delocalization in the two models through the thermally weighted participation ratio.
Section~\ref{sec:conclusion} summarizes the main results of this work.

\section{Models and shared adiabatic geometry}
\label{sec:models}

The ADM describes $N$ identical two-level atoms coupled uniformly to a
single-mode cavity field~\cite{Dicke1954,Buijsman2017}. Setting $\hbar=1$, its Hamiltonian reads
\begin{equation}
\begin{aligned}
\hat{H}_{\ADM}
={}&
\omega \hat{a}^{\dagger}\hat{a}
+
\Delta \hat{J}_{z}
\\
&+
\frac{g_{1}}{\sqrt{N}}
\left(
\hat{a}^{\dagger}\hat{J}_{-}
+
\hat{a}\hat{J}_{+}
\right)
\\
&+
\frac{g_{2}}{\sqrt{N}}
\left(
\hat{a}^{\dagger}\hat{J}_{+}
+
\hat{a}\hat{J}_{-}
\right).
\end{aligned}
\label{eq:H-adm}
\end{equation}
Here $\hat{a}^{\dagger}$ and $\hat{a}$ are the creation and annihilation
operators of the cavity mode with frequency $\omega$,
and $\hat{J}_{z}$ and $\hat{J}_{\pm}$ are the
collective pseudospin operators of the atoms with level splitting $\Delta$. The parameters $g_{1}$ and $g_{2}$ denote the strengths of the rotating
and counterrotating interactions, respectively. Setting $N=1$ in
Eq.~\eqref{eq:H-adm} gives the ARM.
Both Hamiltonians commute with the parity operator
\begin{equation}
\hat{\Pi}
=
\exp
\left[
i\pi
\left(
\hat{a}^{\dagger}\hat{a}
+
\hat{J}_{z}
+
j
\right)
\right],
\label{eq:parity}
\end{equation}
so that each Hilbert space decomposes into even- and odd-parity sectors.
The ARM spectrum can be obtained analytically from parity-resolved transcendental $G$ functions and related constructions~\cite{Braak2011,Xie2014,Xie2017}, and the
model is accordingly classified as integrable by Braak's quantum-number
criterion~\cite{Braak2011,BatchelorZhou2015}.

Along the lines $g_1=0$ and $g_2=0$, the models possess a continuous
$U(1)$ symmetry, with the two integrable axes characterized by different
conserved quantities.
For $g_2=0$, only the rotating-wave interaction
$\hat{a}^{\dagger}\hat{J}_{-}+\hat{a}\hat{J}_{+}$ remains, and the
conserved quantity is the total excitation number $\hat{N}_{+}$.
For $g_1=0$, only the counterrotating interaction
$\hat{a}^{\dagger}\hat{J}_{+}+\hat{a}\hat{J}_{-}$ remains, and the
conserved quantity is instead the excitation-number difference
$\hat{N}_{-}$, where
\begin{equation}
\hat{N}_{\pm}
=
\hat{a}^{\dagger}\hat{a}
\pm
\hat{J}_{z}
+
j.
\label{eq:excitation-number}
\end{equation}
In the ADM, these two integrable axes correspond to the
Tavis--Cummings and anti-Tavis--Cummings models, respectively, with each
$\hat{N}_{\pm}$ multiplet having dimension at most $2j+1$.
Throughout this work, we restrict ourselves to the fully symmetric
sector with $j=N/2$.
For the ARM, the same axes yield the Jaynes--Cummings and
anti-Jaynes--Cummings limits.

In the thermodynamic limit $N\rightarrow\infty$, the ADM undergoes a
superradiant phase transition at the critical coupling
$g_c=\sqrt{\Delta\omega}$~\cite{HeppLieb1973,WangHioe1973}, which gives $g_c=10$ for the parameter used in
our numerical calculations.
The normal phase (NP) occupies $g_1+g_2<g_c$, whereas the superradiant phase
(SP) occupies $g_1+g_2>g_c$.
In the large-frequency-ratio limit $\eta\rightarrow\infty$, the ARM undergoes
the same superradiant transition at $g_1+g_2=g_c$, with the same critical
exponents at finite anisotropy~\cite{Hwang2015,Liu2017}.

\begin{figure}[!tb]
\centering
\includegraphics[width=\columnwidth]{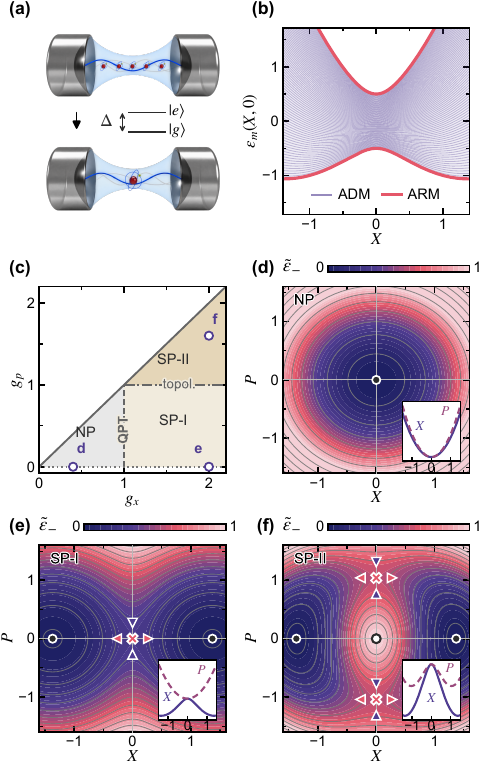}
\caption{Shared adiabatic geometry and inequivalent branch structure. (a) Many-atom ADM (top) and single-atom ARM (bottom). (b) Adiabatic branches along $P=0$ at $(g_x,g_p)=(2,0)$. The thermodynamic ADM fills a continuous purple band, whereas the ARM has only two red branches. (c) Coupling-plane phases: the dashed $g_x=1$ boundary marks the QPT, and the dash-dotted $g_p=1$ boundary marks the stationary-point topology change. (d)--(f) Lowest adiabatic surface in the NP, SP-I, and SP-II, at $(g_x,g_p)=(0.4,0)$, $(2,0)$, and $(2,1.6)$, as marked in (c). Colors show the separately normalized display energy $\tilde{\epsilon}_-$. Filled dark circles, red crosses, and the open circle at the SP-II origin denote minima, saddles, and a maximum, respectively; triangles indicate local downhill directions. Insets show the $P=0$ (solid) and $X=0$ (dashed) cuts. Both models share the lowest-surface stationary-point topology.}
\label{fig:schematic}
\end{figure}

Introducing the
rescaled cavity quadratures
\begin{equation}
X
=
\frac{a^{\dagger}+a}{\sqrt{2\gamma}},
\qquad
P
=
\frac{i(\hat{a}^{\dagger}-\hat{a})}{\sqrt{2\gamma}},
\label{eq:quadratures}
\end{equation}
with $\eta=\Delta/\omega$ and $\gamma=N\eta$, one has $[\hat{X},\hat{P}]=i/\gamma$, so that $X$ and $P$
become classical canonical variables as $\gamma\rightarrow\infty$. Both
$N\rightarrow\infty$ at fixed $\eta$ and
$\eta\rightarrow\infty$ at $N=1$ realize this limit. Dropping the
additive constant $-\omega/2$, the ADM Hamiltonian becomes
\begin{equation}
\hat{H}_{\ADM}
=
\frac{N\Delta}{2}
\left(
X^{2}+P^{2}
\right)
+
\sqrt{\frac{\Delta}{\omega}}\,
\bm{B}_{\rm eff}
\cdot
\bm{J},
\label{eq:H-eff}
\end{equation}
where the collective spin experiences the local effective field
\begin{equation}
\bm{B}_{\rm eff}(X,P)
=
\begin{pmatrix}
\sqrt{2}\,(g_{1}+g_{2})\,X
\\
\sqrt{2}\,(g_{2}-g_{1})\,P
\\
g_{c}
\end{pmatrix}.
\label{eq:Beff}
\end{equation}
Diagonalizing the spin locally along $\bm{B}_{\rm eff}$ yields a family of
adiabatic energy surfaces labeled by the magnetic quantum number
$m=-N/2,\ldots,N/2$,
\begin{equation}
\begin{aligned}
\varepsilon_{m}(X,P)
={}&
\frac{1}{2}
\left(
X^{2}+P^{2}
\right)
\\
&+
\frac{m}{N}
\sqrt{
1
+
2g_{x}^{2}X^{2}
+
2g_{p}^{2}P^{2}
},
\end{aligned}
\label{eq:adiabatic-surfaces}
\end{equation}
where $\varepsilon_{m}=\langle \hat{H}_{\ADM}\rangle/(N\Delta)$ and the
dimensionless couplings are
\begin{equation}
g_{x}
=
\frac{g_{1}+g_{2}}{g_{c}},
\qquad
g_{p}
=
\frac{g_{1}-g_{2}}{g_{c}}.
\label{eq:gxgp}
\end{equation}

The ground-state structure is determined by the lowest adiabatic branch
$m=-N/2$,
\begin{equation}
\begin{aligned}
\varepsilon_{-}(X,P)
={}&
\frac{1}{2}
\left(
X^{2}+P^{2}
\right)
\\
&-
\frac{1}{2}
\sqrt{
1
+
2g_{x}^{2}X^{2}
+
2g_{p}^{2}P^{2}
}.
\end{aligned}
\label{eq:lowest-surface}
\end{equation}
For $g_{1},g_{2}>0$, one has $|g_{x}|>|g_{p}|$, and the stationary-point structure of Eq.~\eqref{eq:lowest-surface} divides the coupling plane into three regions, as illustrated in Fig.~\ref{fig:schematic}(c). In the NP, $1>|g_{x}|>|g_{p}|$, the origin is the unique minimum, as shown in Fig.~\ref{fig:schematic}(d). In SP-I, $|g_{x}|>1>|g_{p}|$, the origin becomes a saddle and two symmetry-related minima emerge along the $X$ axis, as shown in Fig.~\ref{fig:schematic}(e). In SP-II, $|g_{x}|>|g_{p}|>1$, the origin becomes a local maximum and an additional pair of saddles appears along the $P$ axis, as shown in Fig.~\ref{fig:schematic}(f). Across the SP-I--SP-II boundary $|g_p|=1$, the global minima remain at $X_{\pm}$. At fixed $g_x$, the rescaled cavity photon number is independent of anisotropy in the SP, even on the integrable axes.
These nonzero stationary points are located at
\begin{equation}
\begin{aligned}
(X_{\pm},0)
&=
\left(
\pm
\sqrt{
\frac{g_{x}^{4}-1}
{2g_{x}^{2}}
},
0
\right),
\\
(0,P_{\pm})
&=
\left(
0,
\pm
\sqrt{
\frac{g_{p}^{4}-1}
{2g_{p}^{2}}
}
\right),
\end{aligned}
\label{eq:stationary-points}
\end{equation}
valid for $|g_{x}|>1$ and $|g_{p}|>1$, respectively, and their
dimensionless energies, together with that of the origin, are
\begin{equation}
\epsilon_{O}
=
-\frac{1}{2},
\qquad
\epsilon_{\mu}
=
-\frac{1}{4}
\left(
g_{\mu}^{2}
+
g_{\mu}^{-2}
\right),
\label{eq:stationary-energies}
\end{equation}
with $\mu\in\{X,P\}$.
Whenever the excitation energy crosses one of these characteristic values,
the topology of the constant-energy manifold changes, providing the
semiclassical origin of an ESQPT~\cite{Caprio2008,Cejnar2021}.
On the $U(1)$-symmetric lines $g_{x}^{2}=g_{p}^{2}$, the isolated nonzero
stationary points merge into a continuous ring.

\begin{figure*}[!tb]
\centering
\includegraphics[width=\textwidth]{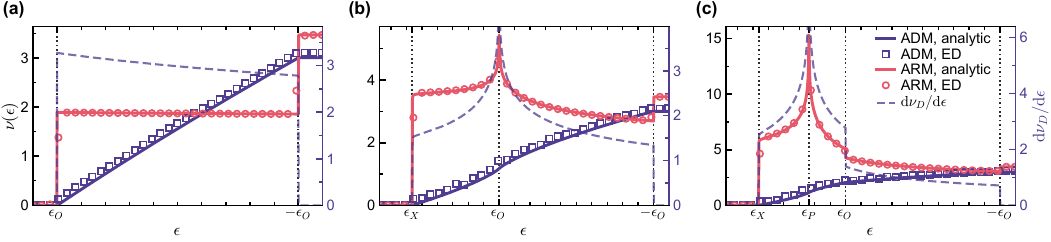}
\caption{Effective densities of states of the ADM and ARM in (a) the NP, (b) SP-I, and (c) SP-II, at $(g_x,g_p)=(0.4,0)$, $(2,0)$, and $(2,1.6)$, respectively. The reduced energy is $\epsilon=E/(N\Delta)$, and the density per unit physical energy is $\nu(\epsilon)=\nu_E(N\Delta\epsilon)=\nu_\epsilon(\epsilon)/(N\Delta)$. Solid curves and open symbols show the analytical results of Appendix~\ref{app:dos} and exact diagonalization, respectively; dashed curves show $d\nu_{\rm D}/d\epsilon$ on the right axis. Vertical dotted lines mark the stationary-point energies of Eq.~\eqref{eq:stationary-energies}. Parameters are $N=30$, $\eta=1$, and $200$ Fock states for the ADM, and $N=1$, $\eta=300$, and $2000$ Fock states for the ARM.}
\label{fig:dos}
\end{figure*}

Although the ground-state structure is shared by the ADM and ARM, the
sets of physically accessible branches are different in the two limits.
For the thermodynamic ADM, the ratio $s=m/N$ becomes continuous and the
spacing between neighboring branches vanishes, so that the adiabatic surfaces
fill the entire band bounded by $\varepsilon_{\pm N/2}$, as illustrated in
Fig.~\ref{fig:schematic}(b).
By contrast, only the two branches $m=\pm 1/2$ exist in the ARM, and in the
large-frequency-ratio limit $\eta\rightarrow\infty$, the separation of the
atomic and photonic time scales suppresses transitions between them, confining
the dynamics to an individual adiabatic branch.

To characterize this difference, we introduce the effective density of states,
\begin{equation}
\nu_{\rm eff}(\epsilon)
=
\frac{1}{2\pi\omega}
\sum_{m=-N/2}^{N/2}
\int
{\rm d}X\,{\rm d}P\,
\delta
\left[
\epsilon-\varepsilon_{m}(X,P)
\right].
\label{eq:dos-definition}
\end{equation}
For the ADM, replacing the sum over $m$ by
$N\int_{-1/2}^{1/2}{\rm d}s$ turns the delta shells into a phase-space
volume bounded by the extremal surfaces, whereas for the ARM the two
delta shells remain.
The density of states can be evaluated analytically in both cases, with
the detailed derivations and resulting expressions given in
Appendix~\ref{app:dos}.

Fig.~\ref{fig:dos} compares the analytical and numerical densities of states
in the three regions.
Overall, the analytical results agree well with exact diagonalization, and the
ADM and ARM become nonanalytic at the same characteristic energies but with
different singularity orders.
At the saddle-point energies $\epsilon_{O}$ in SP-I and $\epsilon_{P}$ in
SP-II, the density of states of the ARM diverges logarithmically, whereas that
of the ADM remains continuous with a divergent first derivative.
At the remaining stationary-point energies, the ARM density of states exhibits
finite discontinuities, while only the first derivative is discontinuous in the
ADM.
Here $-\epsilon_{O}$ denotes the stationary-point energy of the highest
adiabatic branch $m=N/2$.
Thus, although the two limits share a similar static energy-landscape topology,
their densities of states exhibit a clear quantitative distinction.

\section{Inequivalent semiclassical dynamics}
\label{sec:semiclassical}

To reveal the dynamical inequivalence between the two limits, we reconstruct
their semiclassical Hamiltonians.
Combined with the corresponding phase-space equations of motion, the two
limits can be distinguished qualitatively through the evolution of classical
trajectories.
Following the convention of Sec.~\ref{sec:models}, we define the reduced
classical energies as
\begin{equation}
h_{\ADM}
\equiv
\frac{\langle \hat{H}_{\ADM}\rangle}{2N\Delta},
\qquad
h_{\ARM}
\equiv
\frac{\langle \hat{H}_{\ARM}\rangle}{2\Delta},
\label{eq:reduced-energies}
\end{equation}
with $\varepsilon_{m}=2h^{(m)}$ after projection onto an adiabatic branch.
Both quantities are normalized per emitter, allowing the two limits to be
compared directly.

To retain the collective-spin dynamics in the thermodynamic limit
$N\rightarrow\infty$, we represent the collective spin by a Bloch coherent
state~\cite{Radcliffe1971,Arecchi1972},
\begin{equation}
|\beta\rangle
=
\bigl(1+|\beta|^{2}\bigr)^{-N/2}
e^{\beta \hat{J}_{+}}
|N/2,-N/2\rangle ,
\label{eq:bloch}
\end{equation}
with
\begin{equation}
\beta
=
\frac{X_{a}+iP_{a}}
{\sqrt{1-X_{a}^{2}-P_{a}^{2}}},
\qquad
X_{a}^{2}+P_{a}^{2}<1.
\label{eq:bloch-variables}
\end{equation}
After rescaling the cavity quadratures as $X_{c}=X/\sqrt{2}$ and
$P_{c}=P/\sqrt{2}$, we obtain the semiclassical ADM Hamiltonian
\begin{equation}
\begin{aligned}
h_{\ADM}
={}&
\frac{1}{2}
\left(
X_{a}^{2}+P_{a}^{2}-\tfrac{1}{2}
\right)
+
\frac{1}{2}
\left(
X_{c}^{2}+P_{c}^{2}
\right)
\\
&+
S
\sqrt{1-X_{a}^{2}-P_{a}^{2}},
\end{aligned}
\label{eq:h-adm-classical}
\end{equation}
where
\begin{equation}
S
=
g_{x}X_{a}X_{c}
+
g_{p}P_{a}P_{c}.
\label{eq:abbrev}
\end{equation}
Since $2J_{z}/N=2(X_{a}^{2}+P_{a}^{2})-1$, the open disk
$X_{a}^{2}+P_{a}^{2}<1$ covers the Bloch sphere except for the pole
$J_{z}=N/2$, where the projection in Eq.~\eqref{eq:bloch-variables}
becomes singular.

The two canonical pairs $(X_{a},P_{a})$ and $(X_{c},P_{c})$ represent two
independent classical degrees of freedom, so that a trajectory at fixed energy
evolves on a three-dimensional energy shell.
Such an energy shell can support regular islands, chaotic seas, and mixed
phase-space structures~\cite{LichtenbergLieberman1992}.
Defining the phase-space vector
$\bm{z}=(X_a,P_a,X_c,P_c)^{\mathrm T}$, the equations of motion generated by
$h_{\ADM}$ can be written compactly as
\begin{equation}
\dot{\bm{z}}
=
\Omega
\nabla_{\bm{z}} h_{\ADM},
\qquad
\Omega
=
\begin{pmatrix}
0 & 1 & 0 & 0\\
-1 & 0 & 0 & 0\\
0 & 0 & 0 & 1\\
0 & 0 & -1 & 0
\end{pmatrix}.
\label{eq:adm-eom}
\end{equation}

For the ARM in the large-frequency-ratio limit, the dynamics is confined to an
individual adiabatic branch, and we focus on the lower branch.
The corresponding semiclassical Hamiltonian is
\begin{equation}
h_{\ARM}^{-}(X_{c},P_{c})
=
\frac{1}{2}
\left(
X_{c}^{2}+P_{c}^{2}
\right)
-
\frac{1}{4}R_{c},
\label{eq:h-arm-classical}
\end{equation}
with
\begin{equation}
R_{c}
=
\sqrt{
1
+
4g_{x}^{2}X_{c}^{2}
+
4g_{p}^{2}P_{c}^{2}
},
\label{eq:Rc}
\end{equation}
and the equations of motion are
\begin{equation}
\dot{X}_{c}
=
P_{c}
\left(
1-
\frac{g_{p}^{2}}{R_{c}}
\right),
\qquad
\dot{P}_{c}
=
-X_{c}
\left(
1-
\frac{g_{x}^{2}}{R_{c}}
\right).
\label{eq:arm-eom}
\end{equation}
Its trajectories are one-dimensional constant-energy curves in the
$(X_{c},P_{c})$ plane, so no choice of parameters can generate a chaotic sea
on a fixed energy shell.
The resulting single classical degree of freedom arises from the adiabatic
elimination of the qubit in the limit $\eta\rightarrow\infty$, rather than
from the two-dimensional Hilbert space of the qubit itself.

\begin{figure}[!tb]
\centering
\includegraphics[width=\columnwidth]{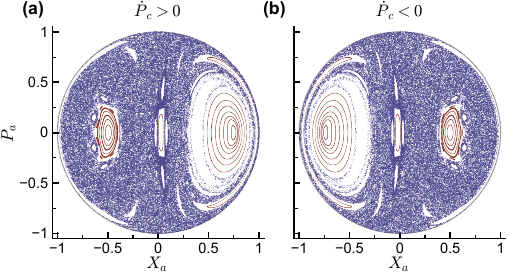}
\caption{Atomic-coordinate Poincar\'e sections of the ADM at $h_D=2.75$, $(g_x,g_p)=(2,1.1)$, and $\eta=1$. Returns to $P_c=0$ are projected onto the $(X_a,P_a)$ plane for (a) $\dot{P}_c>0$ and (b) $\dot{P}_c<0$. Purple and brown points denote selected chaotic and numerically regular trajectories, respectively, classified by finite-time Lyapunov growth. Regular invariant curves coexist with a chaotic sea. The thin gray circle marks the atomic-coordinate boundary $X_a^2+P_a^2=1$.}
\label{fig:poincare}
\end{figure}

To characterize the semiclassical phase-space structure of the ADM, we construct Poincar\'e sections at $h_{\ADM}=2.75$, $(g_x,g_p)=(2,1.1)$, and $\eta=1$. We record the atomic coordinates $(X_a,P_a)$ whenever a trajectory crosses $P_c=0$, separating positive and negative crossings into Fig.~\ref{fig:poincare}(a) and (b), respectively. Using dimensionless time $\tau=\Delta t$, we digscard the transient $\tau\leq t_{\rm tr}=40$ and evaluate the finite-time Lyapunov exponent
\begin{equation}
\lambda_T=\frac{1}{T-t_{\rm tr}}\sum_k\ln\frac{\lVert\delta\bm{z}_k\rVert}{\delta_0},
\label{eq:ftle}
\end{equation}
with the standard tangent-space method~\cite{Benettin1980a,Benettin1980b,Skokos2010}, where $\delta\bm{z}_k$ is the deviation vector before each renormalization and $\delta_0$ is its reset norm. We also evaluate $\lambda_{\rm late}$ over the second half of the post-transient interval. Trajectories are classified as numerically regular when both $|\lambda_T|$ and $|\lambda_{\rm late}|$ are below $0.002$, and as chaotic when $\lambda_T>0.03$ and $\lambda_{\rm late}>0.025$; unresolved cases are omitted. The selected trajectories are integrated to $T=8000$, with eleven chaotic trajectories extended to $T=48000$.

Fig.~\ref{fig:poincare} shows regular islands embedded in a broad chaotic sea. Brown returns trace nested regular curves, whereas purple returns densely populate the chaotic regions. The two crossing directions provide parity-related views of the same energy shell, rather than distinct dynamical regimes. This coexistence illustrates the mixed phase space of the two-degree-of-freedom ADM.

\section{Spectral Statistics of Chaos}
\label{sec:spectral}

Level-spacing statistics distinguish regular spectra, with Poissonian
spacings, from chaotic spectra, which can exhibit random-matrix correlations~\cite{BerryTabor1977,Bohigas1984,Guhr1998}.
For an ordered spectrum $E_n$, we define the
nearest-neighbor spacings $s_n=E_{n+1}-E_n$ and compare their distributions
with the Poisson and GOE forms,
\begin{equation}
P_{\rm P}(s)
=
e^{-s},
\qquad
P_{\rm GOE}(s)
=
\frac{\pi s}{2}
e^{-\pi s^{2}/4}.
\label{eq:reference-distributions}
\end{equation}
A convenient unfolding-free quantity is the adjacent-spacing ratio
~\cite{OganesyanHuse2007,Atas2013},
\begin{equation}
r_n
=
\frac{
\min(s_n,s_{n+1})
}{
\max(s_n,s_{n+1})
}
\in [0,1],
\label{eq:r-ratio}
\end{equation}
whose mean values are
$\overline{r}_{\rm P}=2\ln2-1\simeq0.386$ for Poisson statistics and
$\overline{r}_{\rm GOE}\simeq0.536$ for GOE statistics
~\cite{Atas2013}.
Since both Hamiltonians preserve parity, all statistics are evaluated within
the even-parity sector, and we denote by $\rall$ the average of $r_n$ over the
lowest $10^{3}$ levels.

\begin{figure}[!tb]
\centering
\includegraphics[width=\columnwidth]{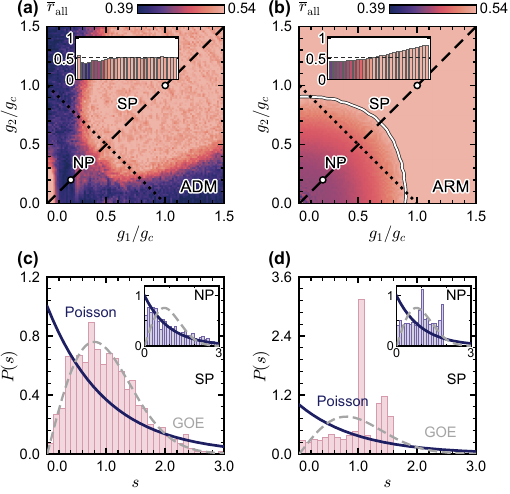}
\caption{Even-parity spectral statistics of the ADM (a,c) and ARM (b,d). (a,b) Mean adjacent-spacing ratio $\overline{r}_{\rm all}$ over the lowest $10^3$ levels. Upper-left insets show its dependence along $g_1=g_2=g$ for $g/g_c=0.05,0.10,\ldots,1.50$, with bar colors following the parent map's clipped scale; horizontal dashed lines mark the GOE reference. The dashed diagonal is $g_1=g_2$, the dotted boundary is $g_1+g_2=g_c$, and the white contour in (b) denotes $\overline{r}_{\rm all}=\overline{r}_{\rm GOE}$. White circles mark the NP and SP samples at $(g_1/g_c,g_2/g_c)=(0.2,0.2)$ and $(1,1)$, respectively, as in Fig.~5. (c,d) Locally unfolded spacing distributions at the SP samples (pale-pink main histograms), with NP samples in the pale-purple upper-right insets. Dark-blue solid and gray dashed curves show the Poisson and GOE references. Parameters are $N=20$, $\eta=1$, and $400$ Fock states for the ADM, and $N=1$, $\eta=200$, and $4000$ Fock states for the ARM, with $g_c=10$.}
\label{fig:spectral}
\end{figure}

Fig.~\ref{fig:spectral}(a) shows a broad nonintegrable region with $\rall$ near the GOE reference, including values above $\overline{r}_{\rm GOE}$. Along the isotropic line $g_1=g_2=g$, the bar inset shows that $\rall$ settles into a stable plateau around $\overline{r}_{\rm GOE}$ over the displayed coupling range. The corresponding SP spacing distribution in Fig.~\ref{fig:spectral}(c) exhibits level repulsion and the broad maximum of the GOE form, confirming the random-matrix-like correlations associated with this plateau. Near the origin, small but nonzero couplings instead produce $\rall\simeq0.9$. At resonance, weak coupling splits the degenerate excitation manifolds of the uncoupled system into nearly equidistant multiplets. Ratios between intramultiplet spacings dominate the average, so these enhanced values reflect regular multiplet structure rather than chaos.

The ARM map in Fig.~\ref{fig:spectral}(b) also contains values at and above $\overline{r}_{\rm GOE}$, but its isotropic-line inset shows a smooth crossing followed by continued growth, rather than a GOE plateau. At the selected SP sample, $\rall\simeq0.654>\overline{r}_{\rm GOE}$, while Fig.~\ref{fig:spectral}(d) shows a sharp peak near $s=1$ superimposed on a broader background. This structure follows from the two adiabatic branches. Below the upper-branch threshold $\epsilon=1/2$, the lower branch produces a regular sequence with smoothly varying spacings, which local unfolding concentrates near $s=1$. Once the upper branch contributes, the interleaving of the two level sequences generates a wider range of spacings. Increasing the coupling while retaining the lowest $10^3$ levels enlarges the fraction of this window below the upper-branch threshold. The growing weight of the regular lower-branch sequence raises $\rall$. In the strong-coupling limit, the displaced minima become nearly harmonic and the fixed-parity spectrum approaches an equidistant oscillator ladder, driving $\rall$ toward unity.

The two models consequently exhibit distinct level-correlation structures. For the ADM, the near-GOE plateau and the corresponding spacing distribution provide consistent spectral signatures of its semiclassical chaotic dynamics. For the ARM, the mean ratio passes through the GOE reference during the evolution toward increasingly regular spacings. The mean adjacent-spacing ratio alone is therefore not a reliable chaos diagnostic for the ARM, and the full distribution $P(s)$ distinguishes a coincident GOE mean from genuine random-matrix correlations.

\section{Thermal Scrambling Correlator}
\label{sec:scrambling}

Thermal OTOCs provide a probe of quantum scrambling in the ADM and ARM~\cite{Swingle2018}.
Within the even-parity sector, the thermal state is defined as
\begin{equation}
\hat{\rho}_{\beta}
=
\frac{
\hat{P}_{+}e^{-\beta \hat{H}}\hat{P}_{+}
}{
Z_{+}
},
\qquad
Z_{+}
=
{\rm Tr}
\left(
\hat{P}_{+}e^{-\beta \hat{H}}
\right),
\label{eq:thermal-state}
\end{equation}
where $\hat{P}_{+}$ projects onto the even-parity sector and
$k_{\rm B}=1$.
The two models are compared at the same dimensionless inverse temperature
$\beta N\Delta=1$, so that the thermal state probes comparable energy regions within their similar static energy-landscape topology.

We probe photon-number scrambling with the unitary operators
\begin{equation}
\hat{W}
=
\hat{V}
=
e^{i\theta\hat{a}^{\dagger}\hat{a}/\gamma},
\qquad
\theta=0.05 ,
\label{eq:B}
\end{equation}
and define the thermal OTOC and scrambling signal as
\begin{equation}
F_{\beta}(t)
=
{\rm Tr}
\left[
\hat{\rho}_{\beta}
\hat{W}^{\dagger}(t)
\hat{V}^{\dagger}
\hat{W}(t)
\hat{V}
\right],
\label{eq:otoc}
\end{equation}
and
\begin{equation}
\Seq(t)
=
1-{\rm Re}\,F_{\beta}(t)
=
\frac{1}{2}
\bigl\langle
[\hat{W}(t),\hat{V}]^{\dagger}
[\hat{W}(t),\hat{V}]
\bigr\rangle_{\beta},
\label{eq:scrambling}
\end{equation}
with
$\hat{W}(t)=e^{i\hat{H}t}\hat{W}e^{-i\hat{H}t}$.
Because the generator is the photon-number operator,
$\Seq(t)$ measures the temporal noncommutativity between
$\hat{a}^{\dagger}\hat{a}(t)$ and $\hat{a}^{\dagger}\hat{a}$, while
$\theta$ sets the perturbation strength.
The probe is parity even and therefore does not couple the two parity sectors.

To expose the origin of the scrambling signal, we represent and diagonalize the
Hamiltonian in the bare basis
$|\alpha\rangle=|n\rangle\otimes|j,m\rangle$.
The energy eigenstates are expanded as
\begin{equation}
|E_{n}\rangle
=
\sum_{\alpha}
c_{\alpha n}
|\alpha\rangle ,
\label{eq:expansion}
\end{equation}
and the probe matrix elements in the energy basis are
\begin{equation}
W_{mn}
=
\sum_{\alpha}
c^{*}_{\alpha m}
c_{\alpha n}
w_{\alpha},
\qquad
w_{\alpha}
=
e^{i\theta n_{\alpha}/\gamma}.
\end{equation}
If every energy eigenstate coincided with a single bare state, $\hat{W}$ would
also be diagonal in the energy basis and hence satisfy
$\hat{W}(t)=\hat{W}$, giving $\Seq(t)=0$.
The scrambling signal can therefore be directly related to the spreading of
energy eigenstates over the bare basis.

This connection becomes explicit in the infinite-time average.
Under the nonresonance condition specified in Appendix~\ref{app:otoc}, the two contributions overlap on the fully
diagonal term, which must be subtracted once, yielding
\begin{equation}
\begin{aligned}
\overline{F_{\beta}}
={}&
\sum_{n} p_{n}
\Bigg[
2\,{\rm Re}
\left(
W_{nn}^{*}
\sum_{k}
W_{kk}|W_{kn}|^{2}
\right)
\\
&\qquad
-
|W_{nn}|^{4}
\Bigg].
\end{aligned}
\label{eq:otoc-average}
\end{equation}
At symmetry-enhanced or accidentally resonant points, additional zero-frequency
terms must be included, while near-degenerate gaps do not modify the strict
infinite-time average but can delay dephasing over a finite observation time.
A detailed derivation of Eq.~\eqref{eq:otoc-average} is given in
Appendix~\ref{app:otoc}.
For generic nonresonant parameters, the long-time plateau is therefore a static
functional of the probe matrix elements in the energy basis.

Eq.~\eqref{eq:B} may also be viewed as a bounded unitary counterpart of
the diagnostic introduced by Buijsman \emph{et al.}~\cite{Buijsman2017}, who
used
$\hat{W}=\hat{V}=\hat{a}^{\dagger}\hat{a}+\mu$
with large $\mu$.
Their construction likewise implements a weak perturbation, but the
nonunitarity of $\hat{V}$ prevents the squared-commutator identity in
Eq.~\eqref{eq:scrambling} from holding.
Expanding their normalized diagnostic at large $\mu$ gives
\begin{equation}
1-
\frac{
{\rm Re}\,F_{\beta}(t)
}{
{\rm Re}\,F_{\beta}(0)
}
=
\frac{4}{\mu^{2}}
\left[
\langle\hat{n}^{2}\rangle_{\beta}
-
{\rm Re}
\langle\hat{n}\hat{n}(t)\rangle_{\beta}
\right]
+
O(\mu^{-3}),
\end{equation}
so that the leading contribution is a two-time correlation function.
Its infinite-time average reduces to
\begin{equation}
\frac{4}{\mu^{2}}
\sum_{n}
p_{n}
{\rm Var}_{n}(\hat{n}),
\end{equation}
the thermally weighted photon-number variance of the energy eigenstates.
This quantity is likewise a static eigenvector functional and is therefore
consistent with the interpretation developed here.

\begin{figure}[!tb]
\centering
\includegraphics[width=\columnwidth]{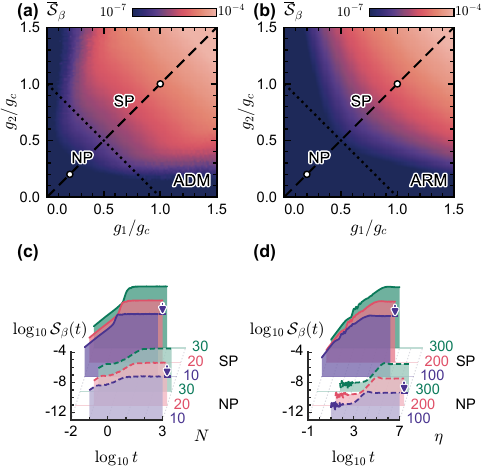}
\caption{Thermal scrambling of the ADM (a,c) and ARM (b,d) at $\beta N\Delta=1$. (a,b) Coupling-plane maps of $\overline{\mathcal{S}}_\beta$ on a shared logarithmic color scale. The dotted boundary is $g_1+g_2=g_c$, and the dashed diagonal is $g_1=g_2$. White circles mark the NP and SP samples at $(g_1/g_c,g_2/g_c)=(0.2,0.2)$ and $(1,1)$, respectively. (c,d) Waterfall traces of $\mathcal{S}_\beta(t)$ at these samples. The ADM traces use $N=10,20,30$ at $\eta=1$, with $200,400,500$ Fock states, respectively; the ARM traces use $\eta=100,200,300$ at $N=1$, with $2000,3000,5000$ Fock states, respectively. Purple, red, and green denote the three conditions in order. Front dashed and back solid curves represent NP and SP. Purple arrows identify the infinite-time predictions for $N=10$ and $\eta=100$ in both phases. Map parameters are $N=10$, $\eta=1$, and $200$ Fock states for the ADM, and $N=1$, $\eta=100$, and $2000$ Fock states for the ARM. The probe is $W=V=\exp(i\theta a^\dagger a/\gamma)$, with $\theta=0.05$, $\gamma=N\Delta/\omega$, and $g_c=10$.}
\label{fig:scrambling}
\end{figure}

Fig.~\ref{fig:scrambling}(a) shows that the ADM long-time scrambling signal $\overline{\Seq}$ is suppressed near the integrable axes and enhanced away from them, with the largest values in the interior of the SP. This pattern is consistent with the GOE-like spectral correlations discussed in the preceding section. The NP and SP traces in Fig.~\ref{fig:scrambling}(c) rise toward and fluctuate around their predicted long-time plateaus. At the marked samples, the SP plateau is approximately $3.2$--$3.5\times10^2$ times the NP plateau, corresponding to about $2.5$ orders of magnitude. Increasing $N$ delays the buildup of the signal, moderately lowers the plateau amplitudes, and reduces late-time fluctuations. The purple arrows mark the NP and SP predictions for $N=10$.

Fig.~\ref{fig:scrambling}(b) shows a similar phase-dependent enhancement in the ARM. The long-time signal remains small in the NP and near the integrable axes, but increases strongly in the interior of the SP. The traces in Fig.~\ref{fig:scrambling}(d) exhibit an SP-to-NP plateau ratio of approximately $2.8$--$2.9\times10^4$, or about $4.45$ orders of magnitude. Increasing $\eta$ shifts the growth and approach to the plateau to later times, while leaving the plateau amplitudes nearly unchanged. The purple arrows identify the corresponding predictions for $\eta=100$. This pronounced scrambling contrast coexists with the non-GOE spacing structure discussed in the preceding section.

A large scrambling signal therefore cannot by itself be regarded as a
signature of quantum chaos.
In the ADM, enhanced scrambling coincides with GOE statistics in the
nonintegrable region, consistent with its chaotic spectral behavior.
By contrast, the ARM exhibits comparably enhanced scrambling even though its
large-frequency-ratio limit lacks the additional classical degree of freedom
required for classical-like chaos.

This separation between chaos and scrambling is not restricted to global
thermal states.
In our companion Letter~\cite{Companion}, we study the FOTOC of localized
wave packets in the ADM.
When a wave packet is prepared at an unstable saddle point, the FOTOC exhibits
early-time exponential growth even when the corresponding energy shell remains
regular.
The growth originates from delocalization driven by the local saddle-point
instability and should therefore be distinguished from chaos of the full energy
shell.
This observation motivates us to ask whether, for global thermal states,
eigenstate delocalization can quantitatively account for the scrambling signal.

\section{Scrambling tracks delocalization}
\label{sec:deloc}

Eigenstate delocalization is quantified by the participation ratio (PR) of
$|E_{n}\rangle$ in the bare parity basis~\cite{Izrailev1990},
\begin{equation}
\PR_{n}
=
\frac{1}{
\sum_{\alpha}
|c_{\alpha n}|^{4}
},
\label{eq:pr}
\end{equation}
and its thermal average is defined using the same ensemble as in
Sec.~\ref{sec:scrambling},
\begin{equation}
\left\langle
\PR
\right\rangle_{\beta}
=
\sum_{n}
p_{n}
\PR_{n},
\qquad
\beta N\Delta=1.
\label{eq:pr-thermal}
\end{equation}
For a state completely localized on a single basis state,
$\PR_{n}=1$, whereas for a state uniformly spread over the full parity sector,
$\PR_{n}=\DP$, where $\DP$ denotes the dimension of that sector.

\begin{figure}[!tb]
\centering
\includegraphics[width=\columnwidth]{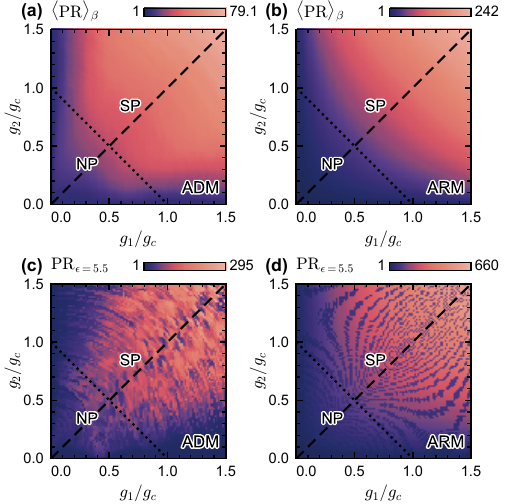}
\caption{Basis delocalization in the ADM (a,c) and ARM (b,d). (a,b) Thermally averaged participation ratio $\langle\mathrm{PR}\rangle_\beta$ at $\beta N\Delta=1$. (c,d) Participation ratio of the even-parity eigenstate nearest the reduced energy $\epsilon=E/(N\Delta)=5.5$. PR is evaluated in the even-parity Fock--spin product basis. The dotted boundary is $g_1+g_2=g_c$, and the dashed diagonal is $g_1=g_2$. Each panel uses its own PR color scale. Parameters are $N=10$, $\eta=1$, and $200$ Fock states for the ADM, and $N=1$, $\eta=100$, and $2000$ Fock states for the ARM, with $g_c=10$.}
\label{fig:deloc}
\end{figure}

Fig.~\ref{fig:deloc}(a,b) shows the thermally averaged participation ratio $\langle\mathrm{PR}\rangle_\beta=\sum_n p_n\mathrm{PR}_n$ at $\beta N\Delta=1$, evaluated in the even-parity bare product basis. In the ADM, the thermal PR remains small near the integrable axes and increases toward the interior of the coupling plane, reaching its largest values in the SP. The ARM exhibits a similar enhancement, with a more pronounced contrast between the NP and SP. These distributions closely resemble the long-time scrambling maps in Fig.~\ref{fig:scrambling}(a,b), associating the enhanced thermal plateau with the spreading of thermally populated eigenstates over the bare basis.

To compare eigenstate delocalization at a common reduced energy, we select the even-parity eigenstate nearest $\epsilon=E/(N\Delta)=5.5$ at each coupling point. The resulting PR maps in Fig.~\ref{fig:deloc}(c,d) show substantial delocalization away from the integrable axes in both models. Compared with the thermal averages, these single-eigenstate maps exhibit more pronounced local structure. The ADM eigenstates are strongly delocalized in the GOE-like region of Fig.~\ref{fig:spectral}(a), linking quantum chaos with bare-basis delocalization. The ARM eigenstates also show strong delocalization with bandlike PR patterns, despite non-GOE spacing statistics.

The discussion above focuses on thermally populated and highly excited
eigenstates, which are directly relevant to the thermal scrambling and spectral
diagnostics considered here.
For completeness, the ground-state delocalization of the ADM and ARM is
examined separately in Appendix~\ref{app:groundstate} through the participation
ratio, Shannon entropy, and Husimi distribution.
These ground-state measures mainly reflect the restructuring of the
superradiant energy landscape and should therefore be distinguished from the
delocalization associated with quantum chaos at higher energies.

\section{Conclusion}
\label{sec:conclusion}

This work investigates the relation between quantum chaos and scrambling in
the ADM and ARM and shows that the two signatures need not coincide.
Although the thermodynamic ADM and the large-frequency-ratio ARM share the
same family of adiabatic energy surfaces, their semiclassical limits retain
different numbers of classical degrees of freedom.
This distinction is reflected quantitatively in the different ESQPT
singularities of their densities of states and qualitatively in their
semiclassical dynamics.
The ADM supports chaotic energy shells, whereas the
reduced ARM is restricted to one-dimensional constant-energy trajectories.

The spectral and scrambling diagnostics reveal the consequence of this
difference.
The ADM develops GOE level statistics in the nonintegrable region, while the
ARM shows no corresponding GOE correlations.
In contrast, the thermal OTOC is strongly enhanced in both models, showing
that substantial scrambling can occur in the absence of classically chaotic
dynamics.

The thermally weighted participation ratio further shows that the scrambling
signal in both models closely tracks eigenstate delocalization in the bare
parity basis.
The ARM therefore provides a direct example in which strong delocalization and
scrambling occur without GOE statistics or classical-like chaos.
These results indicate that eigenstate delocalization alone is not sufficient
for quantum chaos, and that identifying chaos in quantum systems requires its
combination with the classical degrees of freedom retained in the corresponding
semiclassical limit.

\section*{Data availability}

The data that support the findings of this study, and the scripts that generate
the figures from them, are available from the corresponding author upon
reasonable request.

\begin{acknowledgments}
This work was supported by the National Natural Science Foundation of China
(Grant Nos.~12205383 and 12274470) and the Hunan Provincial Natural Science
Foundation (Grant No.~2024JJ6483).
\end{acknowledgments}

\appendix

\section{Analytical effective densities of states}
\label{app:dos}

Here we evaluate Eq.~\eqref{eq:dos-definition} for both models. For the ADM in
the thermodynamic limit the branch label becomes continuous,
$\sum_{m}\rightarrow N\!\int_{-1/2}^{1/2}{\rm d}s$ with $s=m/N$, and performing
the integral over $s$ gives
\begin{equation}
\begin{aligned}
\nu_{\ADM}(\epsilon)
={}&
\frac{N}{2\pi\omega}
\!\int\!
\frac{{\rm d}X\,{\rm d}P}{R(X,P)}
\,
\Theta
\!
\left[
\varepsilon_{N/2}-\epsilon
\right]
\\
&\times
\Theta
\!
\left[
\epsilon-\varepsilon_{-N/2}
\right],
\end{aligned}
\label{eq:app-dos-adm}
\end{equation}
where
\begin{equation}
R(X,P)
=
\sqrt{
1
+
2g_{x}^{2}X^{2}
+
2g_{p}^{2}P^{2}
}.
\label{eq:app-R}
\end{equation}
For the ARM only the two branches survive and
\begin{equation}
\nu_{\ARM}(\epsilon)
=
\frac{1}{2\pi\omega}
\!\int\!
{\rm d}X\,{\rm d}P
\sum_{\sigma=\pm1}
\delta
\!
\left[
\varepsilon_{\sigma/2}
-
\epsilon
\right].
\label{eq:app-dos-arm}
\end{equation}

The remaining integrations are performed in polar coordinates
$X=r\cos\theta$, $P=r\sin\theta$ with $\rho=r^{2}$. Introducing
\begin{equation}
\Phi(\theta)
=
2g_{x}^{2}\cos^{2}\theta
+
2g_{p}^{2}\sin^{2}\theta
\label{eq:app-Phi}
\end{equation}
and
\begin{equation}
D_{\theta}
=
\sqrt{
\frac{\Phi^{2}}{16}
+
\frac{\epsilon\Phi}{2}
+
\frac{1}{4}
},
\label{eq:app-D}
\end{equation}
the radial integrals can be done in closed form and leave the angular kernels
\begin{equation}
\begin{aligned}
f_{\ADM}^{(0)}
&=
\frac{1}{2}
+
\frac{2D_{\theta}-1}{\Phi},
&
f_{\ADM}^{(-)}
&=
\frac{4D_{\theta}}{\Phi},
\\[2pt]
f_{\ARM}^{(0)}
&=
1
+
\frac{\Phi}{4D_{\theta}},
&
f_{\ARM}^{(-)}
&=
\frac{\Phi}{2D_{\theta}},
\end{aligned}
\label{eq:app-kernels}
\end{equation}
whose values at $\Phi=0$ are understood as continuous limits. Denoting their
angular integrals by
\begin{equation}
\mathcal{I}_{\alpha}^{(0)}
=
\!\int_{0}^{2\pi}\!
f_{\alpha}^{(0)}\,{\rm d}\theta,
\qquad
\mathcal{I}_{\alpha}^{(-)}
=
\!\int_{\mathcal{A}_{\epsilon}}\!
f_{\alpha}^{(-)}\,{\rm d}\theta,
\label{eq:app-integrals}
\end{equation}
with $\alpha\in\{\ADM,\ARM\}$, the densities of states are
\begin{equation}
\nu_{\ADM}(\epsilon)
=
\frac{N}{2\pi\omega}
\begin{cases}
2\pi,
&
\epsilon>-\epsilon_{O},
\\[2pt]
\mathcal{I}_{\ADM}^{(0)},
&
\epsilon_{O}<\epsilon\leq-\epsilon_{O},
\\[2pt]
\mathcal{I}_{\ADM}^{(-)},
&
\epsilon\leq\epsilon_{O},
\end{cases}
\label{eq:app-nu-adm}
\end{equation}
and
\begin{equation}
\nu_{\ARM}(\epsilon)
=
\frac{1}{2\pi\omega}
\begin{cases}
4\pi,
&
\epsilon>-\epsilon_{O},
\\[2pt]
\mathcal{I}_{\ARM}^{(0)},
&
\epsilon_{O}<\epsilon\leq-\epsilon_{O},
\\[2pt]
\mathcal{I}_{\ARM}^{(-)},
&
\epsilon\leq\epsilon_{O}.
\end{cases}
\label{eq:app-nu-arm}
\end{equation}
The allowed angular domain in the low-energy regime is
$\mathcal{A}_{\epsilon}=\{\theta:\Phi(\theta)\geq\Phi_{c}\}$ with
\begin{equation}
\Phi_{c}(\epsilon)
=
-4\epsilon
+
2\sqrt{4\epsilon^{2}-1},
\qquad
\epsilon\leq\epsilon_{O}.
\label{eq:app-Phic}
\end{equation}
The three phases are thus described by the same kernels, and are distinguished
only by whether $\mathcal{A}_{\epsilon}$ is empty, complete, or restricted to
a subset of angles: empty for $\epsilon\leq\epsilon_{O}$ in the NP,
restricted for $\epsilon_{X}<\epsilon\leq\epsilon_{O}$ in SP-I, and
complete for $\epsilon_{P}<\epsilon\leq\epsilon_{O}$ but restricted for
$\epsilon_{X}<\epsilon\leq\epsilon_{P}$ in SP-II.

\section{Infinite-time average of the thermal OTOC}
\label{app:otoc}

\begin{figure*}[t]
\centering
\includegraphics[width=\textwidth]{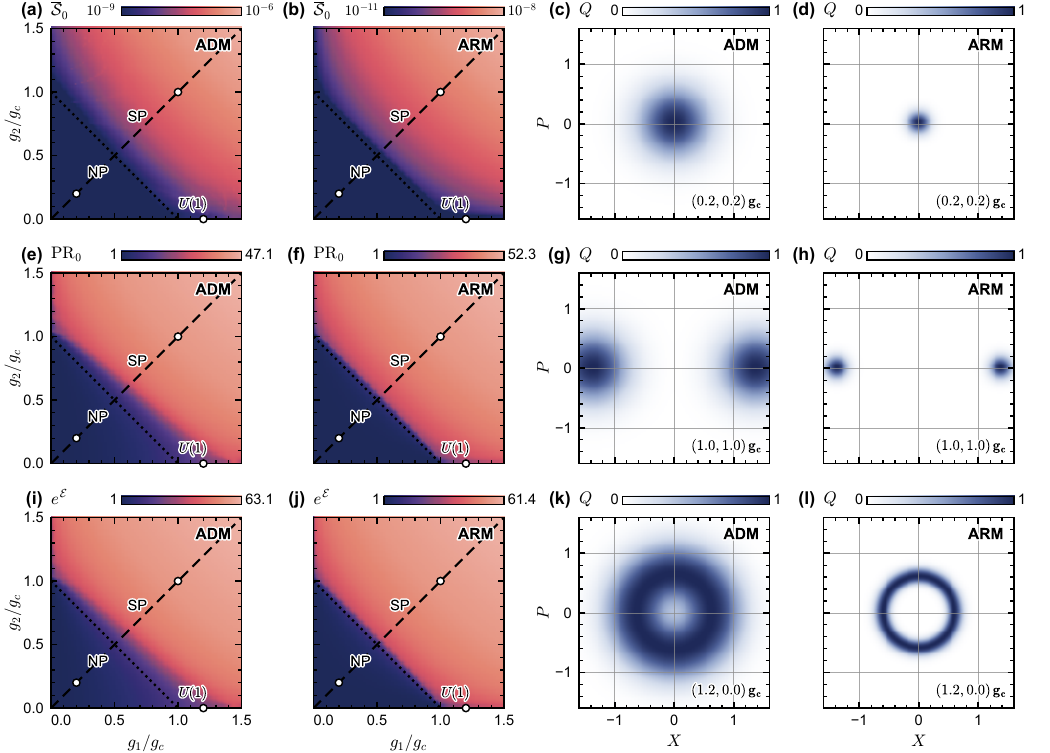}
\caption{
Ground-state delocalization of the ADM and ARM. Panels (a,b), (e,f), and (i,j) show $\overline{\mathcal{S}}_0$, $\mathrm{PR}_0$, and $e^{\mathcal{E}}$, respectively. Panels (c,d), (g,h), and (k,l) show the corresponding Husimi $Q$ functions at the NP, SP, and $U(1)$ points. Parameters are $N=10$, $\eta=1$, and $200$ Fock states for the ADM, and $N=1$, $\eta=100$, and $2000$ Fock states for the ARM, with $g_c=10$.
}
\label{fig:groundstate}
\end{figure*}

Here we derive the infinite-time average of the thermal OTOC used in
Sec.~\ref{sec:scrambling}. Within the chosen parity sector, the thermal
density matrix is diagonal in the energy eigenbasis,
\begin{equation}
\hat{\rho}_{\beta}
=
\sum_{n} p_{n}
|E_{n}\rangle\langle E_{n}|,
\qquad
p_{n}
=
\frac{e^{-\beta E_{n}}}{Z_{+}}.
\label{eq:app-thermal}
\end{equation}
Since $\hat{W}=\hat{V}$ in Eq.~\eqref{eq:B}, we denote the matrix elements of
the scrambling operator in the energy basis by
\begin{equation}
W_{mn}
=
\langle E_{m}|\hat{W}|E_{n}\rangle.
\label{eq:app-Wmn}
\end{equation}
Its time-dependent matrix elements are therefore
\begin{equation}
\langle E_{m}|\hat{W}(t)|E_{n}\rangle
=
e^{i(E_{m}-E_{n})t}W_{mn}.
\label{eq:app-Wt}
\end{equation}

Starting from Eq.~\eqref{eq:otoc},
\begin{equation}
F_{\beta}(t)
=
{\rm Tr}
\left[
\hat{\rho}_{\beta}
\hat{W}^{\dagger}(t)
\hat{W}^{\dagger}
\hat{W}(t)
\hat{W}
\right],
\label{eq:app-F-start}
\end{equation}
we insert three resolutions of the identity in the energy basis. This gives
\begin{equation}
\begin{aligned}
F_{\beta}(t)
={}&
\sum_{n,k,l,m}
p_{n}\,
W_{kn}^{*}
W_{lk}^{*}
W_{lm}
W_{mn}
\\
&\times
e^{i(E_{n}-E_{k}+E_{l}-E_{m})t}.
\end{aligned}
\label{eq:app-F-expand}
\end{equation}
For convenience, we define the corresponding frequency difference as
\begin{equation}
\Omega_{nklm}
=
E_{n}-E_{k}+E_{l}-E_{m}.
\label{eq:app-frequency}
\end{equation}
The finite-time average of each oscillatory factor is
\begin{equation}
\begin{aligned}
\frac{1}{T}
\int_{0}^{T}
e^{i\Omega t}\,{\rm d}t
={}&
e^{i\Omega T/2}
\frac{\sin(\Omega T/2)}
{\Omega T/2}.
\end{aligned}
\label{eq:app-dephasing}
\end{equation}
Consequently, in the limit $T\rightarrow\infty$, all terms with
$\Omega_{nklm}\neq0$ dephase, and only the exact zero-frequency condition
\begin{equation}
E_{n}-E_{k}
=
E_{m}-E_{l}
\label{eq:app-zero-frequency}
\end{equation}
contributes to the infinite-time average.

For generic parameters within a fixed symmetry sector, we impose the
nonresonance condition that Eq.~\eqref{eq:app-zero-frequency} admits only the
two trivial pairings
\begin{equation}
k=n,\quad m=l,
\qquad\text{or}\qquad
l=k,\quad m=n.
\label{eq:app-pairings}
\end{equation}
For the first pairing, $k=n$ and $m=l$, the surviving contribution is
\begin{equation}
\begin{aligned}
F_{\beta}^{(1)}
={}&
\sum_{n,l}
p_{n}
W_{nn}^{*}
W_{ln}^{*}
W_{ll}
W_{ln}
\\
={}&
\sum_{n}
p_{n}
W_{nn}^{*}
\sum_{l}
W_{ll}|W_{ln}|^{2}.
\end{aligned}
\label{eq:app-pairing1}
\end{equation}
For the second pairing, $l=k$ and $m=n$, we obtain
\begin{equation}
\begin{aligned}
F_{\beta}^{(2)}
={}&
\sum_{n,k}
p_{n}
W_{kn}^{*}
W_{kk}^{*}
W_{kn}
W_{nn}
\\
={}&
\sum_{n}
p_{n}
W_{nn}
\sum_{k}
W_{kk}^{*}|W_{kn}|^{2}.
\end{aligned}
\label{eq:app-pairing2}
\end{equation}
The fully diagonal terms $n=k=l=m$ satisfy both pairings simultaneously and
are therefore counted twice. Their contribution is
\begin{equation}
F_{\beta}^{(0)}
=
\sum_{n}
p_{n}|W_{nn}|^{4},
\label{eq:app-overlap}
\end{equation}
which must be subtracted once.

Combining the two pairings and removing their overlap gives
\begin{equation}
\begin{aligned}
\overline{F_{\beta}}
={}&
2\sum_{n}p_{n}\,
{\rm Re}
\left(
W_{nn}^{*}
\sum_{k}
W_{kk}|W_{kn}|^{2}
\right)
\\
&-
\sum_{n}p_{n}|W_{nn}|^{4}.
\end{aligned}
\label{eq:app-otoc-average}
\end{equation}
Equivalently,
\begin{equation}
\begin{aligned}
\overline{F_{\beta}}
={}&
\sum_{n}p_{n}
\Bigg[
2\,{\rm Re}
\left(
W_{nn}^{*}
\sum_{k}W_{kk}|W_{kn}|^{2}
\right)
\\
&\hspace{3.5em}
-
|W_{nn}|^{4}
\Bigg],
\end{aligned}
\label{eq:app-otoc-average-compact}
\end{equation}
which is the result quoted in Eq.~\eqref{eq:otoc-average}.

\section{Ground-state delocalization}
\label{app:groundstate}

The ground-state delocalization is further characterized by the Shannon
entropy,
\begin{equation}
\mathcal{E}
=
-\sum_{\alpha}\vert c_{\alpha 0}\vert^{2}\ln \vert c_{\alpha 0}\vert^{2}.
\label{eq:ground-entropy}
\end{equation}
The quantity $e^{\mathcal{E}}$ gives an effective number of occupied basis
states and, together with $\mathrm{PR}_{0}$, provides a complementary measure
of ground-state delocalization.

As shown in Fig.~\ref{fig:groundstate}, $\overline{\mathcal{S}}_{0}$,
$\mathrm{PR}_{0}$, and $e^{\mathcal{E}}$ exhibit similar structures in both
models.
The corresponding Husimi distributions~\cite{Husimi1940} show that the ground state is localized
near the origin in the NP, splits into two symmetry-related lobes in the SP,
and forms a ring along the $U(1)$-symmetric axis.
These features indicate that the ground-state delocalization mainly reflects
the restructuring of the superradiant energy landscape and should not by
itself be interpreted as a signature of quantum chaos.

\end{document}